\documentclass[conference]{ieeeconf}
\IEEEoverridecommandlockouts                              
\usepackage{graphicx}
\usepackage{amsmath,amssymb,amsfonts}
\usepackage{algorithmic}
\usepackage{graphicx}
\usepackage{algorithm,algorithmic}
\usepackage{booktabs}
\usepackage{multirow}
\usepackage{cite}
\usepackage[colorlinks,urlcolor=blue,linkcolor=blue,citecolor=blue]{hyperref}

\title{Learning to Select Like Humans: Explainable Active Learning for Medical Imaging}

\author{Ifrat Ikhtear Uddin$^{1}$, Longwei Wang$^{1}$, Xiao Qin$^{2}$, Yang Zhou$^{2}$ and KC Santosh$^{1}$
\thanks{$^{1}$Ifrat Ikhtear Uddin, Longwei Wang, KC Santosh are with AI Research, Department of Computer Science, University of South Dakota, Vermillion, SD 57069, USA
{\tt\small ifratikhtear.uddin@coyotes.usd.edu},
{\tt\small longwei.wang@usd.edu},
{\tt\small kc.santosh@usd.edu}}
\thanks{$^{2}$Xiao Qin and Yang Zhou  with Department of Computer Science and Software Engineering, Auburn University, Auburn, AL 36849, USA
{\tt\small xqin@auburn.edu}, {\tt\small yangzhou@auburn.edu}}}

\begin{document}

\maketitle
\thispagestyle{empty}
\pagestyle{empty}


\begin{abstract}

Medical image analysis requires substantial labeled data for model training, yet expert annotation is expensive and time-consuming. Active learning (AL) addresses this challenge by strategically selecting the most informative samples for the annotation purpose, but traditional methods solely rely on predictive uncertainty while ignoring whether models learn from clinically meaningful features a critical requirement for clinical deployment.
We propose an explainability-guided active learning framework that integrates spatial attention alignment into a sample acquisition process. Our approach advocates for a dual-criterion selection strategy combining: (i) classification uncertainty to identify informative examples, and (ii) attention misalignment with radiologist-defined regions-of-interest (ROIs) to target samples where the model focuses on incorrect features. By measuring misalignment between Grad-CAM attention maps and expert annotations using \emph{Dice similarity}, our acquisition function judiciously identifies samples that enhance both predictive performance and spatial interpretability.
We evaluate the framework using three expert-annotated medical imaging datasets, namely, BraTS (MRI brain tumors), VinDr-CXR (chest X-rays), and SIIM-COVID-19 (chest X-rays). Using only 570 strategically selected samples, our explainability-guided approach consistently outperforms random sampling across all the datasets, achieving 77.22\% accuracy on BraTS, 52.37\% on VinDr-CXR, and 52.66\% on SIIM-COVID. Grad-CAM visualizations confirm that the models trained by our dual-criterion selection focus on diagnostically relevant regions, demonstrating that incorporating explanation guidance into sample acquisition yields superior data efficiency while maintaining clinical interpretability.

\end{abstract}

\section{INTRODUCTION}

Active learning (AL) has emerged as a principled solution to the annotation bottleneck in medical image analysis, where acquiring expert labels for MRI, CT, and chest X-ray data is time-consuming, expensive, and requires specialized clinical knowledge~\cite{he2016deep, chen2025survey, wang2019representation, 10928239,10928056,10927825}. By strategically selecting which samples to annotate, AL maximizes model improvement per labeling effort~\cite{ren2021surveydeepactivelearning, aghdam2019active} making it especially valuable in data-scarce clinical settings~\cite{ranabhat2025multi, wang2025bridging}. Yet despite decades of progress, a fundamental question in AL design remains unresolved: \textbf{how should we decide which samples are worth annotating?}

\begin{figure}
    \centering
    \includegraphics[width=.95\linewidth]{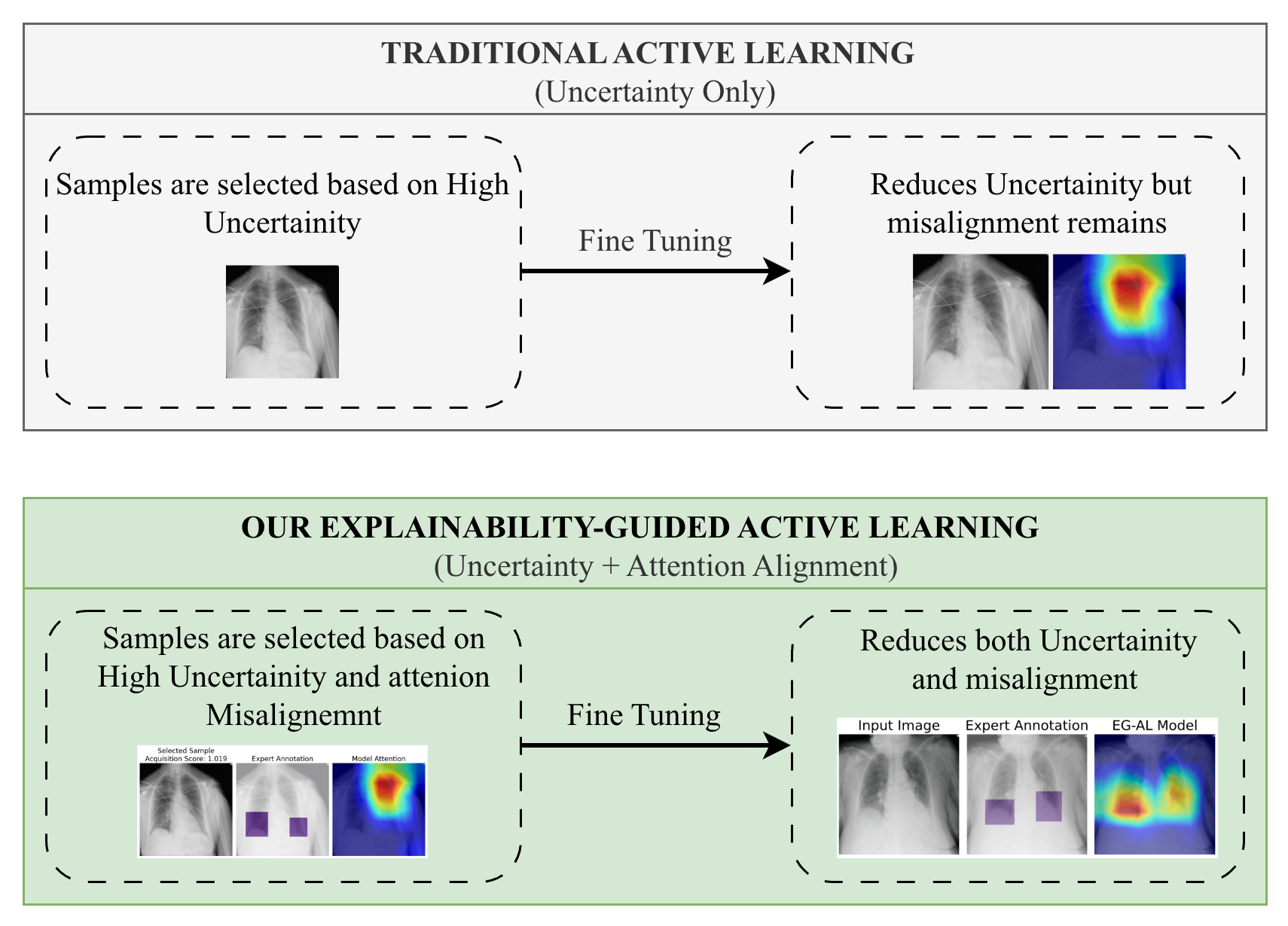}
    \caption{Traditional AL only considers uncertainty. It misses samples where the model is confident but focuses on the wrong features. Our dual-criterion approach catches both failure modes.}
    \label{fig:motivation}
\vspace{-5mm}
\end{figure}

Existing methods answer this almost exclusively through predictive uncertainty entropy~\cite{wang2017cost} or ensemble disagreement. But uncertainty captures only one mode of model failure. A model may confidently predict the correct class while attending to clinically irrelevant regions a failure that is invisible to any uncertainty-based criterion, yet critical for clinical deployment~\cite{wang2019layer, wang2024dense}. Neural networks are well known to exploit spurious correlations~\cite{zhang2025uncertainty}, and uncertainty-based selection has no mechanism to detect this. XAI methods such as Grad-CAM~\cite{selvaraju2017grad} and Winsor-CAM~\cite{wall2025winsor} make spatial attention measurable at inference time, and prior work has used this signal to improve \textit{how} models train~\cite{vsefvcik2023improving, caragliano2025doctorintheloopexplainablemultiviewdeep, sun2021explanation, wang2025explainability, uddin2025expert} but no acquisition function has used it to guide \textit{which} samples to annotate.

\begin{figure*}[htb]
    \centering
    \includegraphics[width=0.8\textwidth]{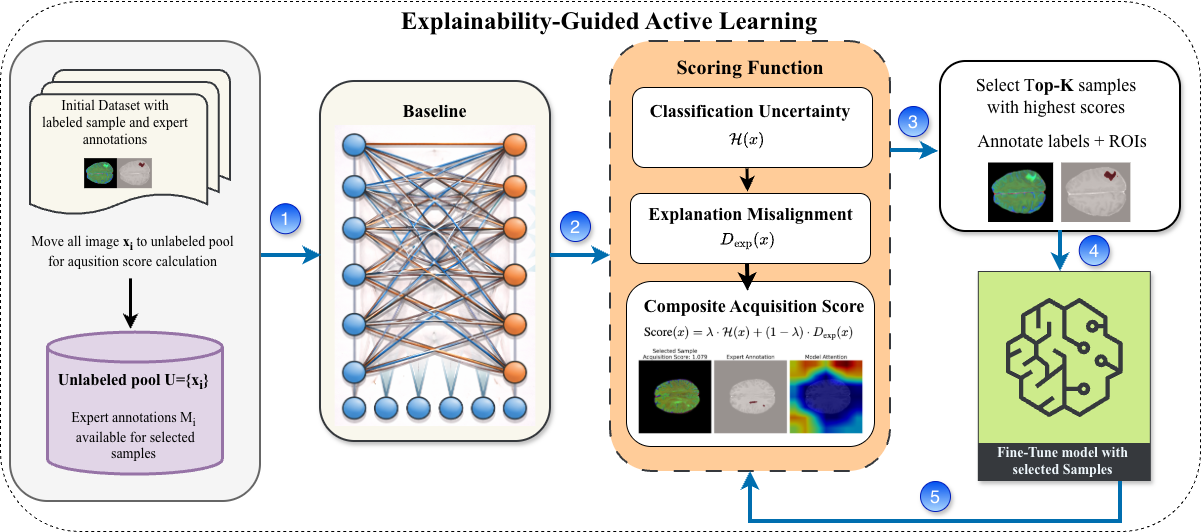}
    \caption{EG-AL framework operating in an iterative cycle: (1) unlabeled 
    pool $U=\{x_i\}$ with expert annotations; (2) baseline model trained on 
    a small labeled seed set to compute acquisition scores; (3) composite 
    scoring combining classification uncertainty $\mathcal{H}(x)$ and 
    explanation misalignment $D_{\text{exp}}(x)$; (4) top-K sample selection 
    with expert annotation of labels and ROI masks; (5) model fine-tuning 
    with explanation-guided supervision, feeding back into the next iteration.}
    \label{fig:active-learning-frm}
    \vspace{-5mm}
\end{figure*}

We close this gap by asking: \textbf{can attention alignment with expert annotations guide sample selection to improve data efficiency beyond what uncertainty alone achieves?} We argue the answer is yes, and that effective acquisition requires dual-criterion reasoning  jointly considering \textbf{what the model does not know} (high uncertainty) and \textbf{where the model is wrong} (attention misalignment with expert ROIs). As illustrated in Fig.~\ref{fig:motivation}, uncertainty-only selection systematically misses the second failure mode, while our approach catches both. To this end, we propose \textbf{Explainability-Guided Active Learning (EG-AL)}, a sample acquisition framework that introduces a composite scoring function combining classification entropy with Grad-CAM-based attention misalignment measured via Dice similarity against expert-annotated ROIs. The result is an acquisition criterion that identifies samples informative along two orthogonal dimensions: epistemic uncertainty about the label, and spatial misalignment between model attention and clinical ground truth. Our main contributions are:

\begin{enumerate}
    \item A novel dual-criterion acquisition function that, for the first time, incorporates spatial explanation misalignment  measured via Dice similarity between Grad-CAM attention maps and expert ROIs  directly into the AL sample selection process.
    \item A formal characterization of three distinct failure patterns captured by the acquisition score: high uncertainty with high misalignment, high uncertainty with low misalignment, and the clinically critical case of low uncertainty with high misalignment that uncertainty-only methods systematically miss.
    \item Empirical validation across three expert-annotated medical imaging datasets and two modalities (MRI, X-ray), consistently outperforming random sampling with only 570 strategically selected samples and demonstrating that explanation quality is a viable and complementary signal to uncertainty in AL acquisition.
\end{enumerate}

\section{Related Work}

\subsection{Uncertainty-Based Acquisition in Active Learning}
AL reduces annotation cost by querying the most informative unlabeled samples~\cite{hsu2015active, ho2024learning, ravi2017optimization}, making it especially valuable in medical imaging where expert annotators are scarce~\cite{biswas2023active, gaillochet2023active}. Dominant acquisition strategies rely on predictive uncertainty  entropy, margin sampling, or ensemble disagreement~\cite{wang2017cost}  to proxy sample informativeness. Methods such as ACFT~\cite{zhou2021active, zhan2025new} and query-by-committee approaches~\cite{hao2021transfer} have demonstrated strong annotation efficiency across clinical tasks. However, all these criteria share a common blind spot: they measure uncertainty about the \textit{label}, not about \textit{where} the model is attending. A confidently mispredicted sample  one where the model attends to spurious regions  scores low on every uncertainty criterion and is never selected, despite representing a high-value annotation opportunity.

\subsection{Spatial Attention as a Diagnostic Signal}
XAI methods such as Grad-CAM~\cite{selvaraju2017grad} and SHAP~\cite{lundberg2017unified} expose which image regions drive model predictions, and have been widely applied in medical tasks including retinopathy detection, chest X-ray diagnosis, and tumor segmentation~\cite{loh2022application, sadeghi2024review, wang2021explaining, wang2024enhanced}. Critically, attention maps provide a measurable signal of \textit{spatial reasoning quality}  the degree to which a model's focus aligns with clinically meaningful regions  that is available at inference time without requiring ground-truth labels~\cite{wang2021improving, xiao2022looking, shi2019deep, wang2014congestion}. This positions attention misalignment not merely as a post-hoc interpretability tool, but as an actionable criterion for identifying model failure modes that uncertainty metrics cannot detect.

\subsection{Explanation-Guided Learning vs. Explanation-Guided Selection}
Prior work has integrated explanation signals into model \textit{training}  using LRP to supervise attention toward glioma regions~\cite{vsefvcik2023improving}, radiologist ROIs to guide focus in lung cancer CT~\cite{caragliano2025doctorintheloopexplainablemultiviewdeep}, and explanation-guided objectives in few-shot and cross-domain settings~\cite{sun2021explanation, uddin2025expert}. These approaches demonstrate that spatial supervision improves both accuracy and interpretability. However, they all presuppose a fixed labeled set and address \textit{how} to train on it  none use explanation quality to determine \textit{which} samples to annotate in the first place. This distinction is the core gap our work addresses: we introduce spatial attention misalignment as an acquisition criterion, enabling AL to select samples that correct not only decision boundaries but also the spatial reasoning underlying model predictions.

\section{Methodology}

\subsection{Problem Formulation}


We consider a pool $\mathcal{U} = \{x_i\}_{i=1}^{M}$ where class labels 
$y_i \in \{1, \ldots, N\}$ are withheld and revealed only upon selection, 
while spatial ROI masks $\text{ESM}(x_i)$  indicating diagnostically 
relevant regions such as tumor boundaries in BraTS or bounding boxes 
in VinDr-CXR and SIIM-COVID  are available pool-wide, as they are 
required to compute $D_{\text{exp}}$ during acquisition 
scoring~\cite{zhou2021active, hao2021transfer}. Starting from a small 
labeled seed set $\mathcal{S}_0$, our goal is to iteratively select 
batches from $\mathcal{U}$ that, when added to the training set and used 
for model retraining, maximize both classification accuracy and spatial 
interpretability. Effective acquisition 
must capture two orthogonal failure modes  label uncertainty and spatial 
misalignment  neither of which subsumes the other: a sample can be 
uncertain yet spatially correct, or confidently predicted yet attending 
to entirely wrong regions. An acquisition function blind to the second 
failure mode will systematically miss samples that are critical for 
correcting the model's spatial reasoning, regardless of how well it 
identifies uncertain ones.

\subsection{Dual-Criterion Acquisition Function}

For each $x \in \mathcal{U}$, we compute an acquisition score balancing two complementary criteria.

\subsubsection{Classification Uncertainty}
We quantify predictive uncertainty via Shannon entropy:
\begin{equation}
\mathcal{H}(x) = - \sum_{k=1}^{N} p(y=k|x) \log p(y=k|x),
\label{eq:entropy}
\end{equation}
where $p(y=k|x)$ is the predicted class probability. High entropy identifies samples near decision boundaries where label ambiguity is high.

\subsubsection{Explanation Misalignment}
To detect spatially wrong predictions, we use Grad-CAM~\cite{selvaraju2017grad} to generate an attention map $\text{CAM}_{\hat{y}}(x)$ for the predicted class $\hat{y} = \arg\max_k\, p(y=k|x)$, and measure its divergence from the expert annotation $\text{ESM}(x)$ via Dice distance:
\begin{equation}
D_{\text{exp}}(x) = 1 - \frac{2 \cdot |\text{CAM}_{\hat{y}}(x) \cap \text{ESM}(x)|}{|\text{CAM}_{\hat{y}}(x)| + |\text{ESM}(x)|},
\label{eq:dice_misalignment}
\end{equation}
where $|\cdot|$ denotes pixel-sum for normalized heatmaps or cardinality for binary masks, and $\cap$ is element-wise multiplication. A high $D_{\text{exp}}(x)$ indicates the model is attending to clinically irrelevant regions  a failure mode invisible to uncertainty metrics. Dice is chosen specifically because it handles the spatial imbalance between compact expert ROIs and diffuse attention maps without requiring a hard threshold.

\subsubsection{Composite Acquisition Score}
Both criteria are combined into a single score:
\begin{equation}
\text{Score}(x) = \lambda \cdot \mathcal{H}(x) + (1 - \lambda) \cdot D_{\text{exp}}(x),
\label{eq:composite_score}
\end{equation}
where $\lambda \in [0,1]$ governs the uncertainty-misalignment trade-off. Setting $\lambda=1$ recovers standard entropy-based AL; $\lambda=0$ selects purely on spatial misalignment. The composite score identifies three distinct failure patterns: \textbf{high uncertainty, high misalignment} (improves both classification and attention), \textbf{high uncertainty, low misalignment} (refines decision boundaries), and \textbf{low uncertainty, high misalignment}  the clinically critical case where the model is confidently wrong in its spatial reasoning, which uncertainty-only methods cannot detect. We select $\lambda$ via grid search; $\lambda=0.5$ achieves strong performance across datasets.

\subsection{Iterative Acquisition Procedure}

\begin{algorithm}[tbp]
\caption{Explainability-Guided Active Learning}
\label{alg:active_learning}
\begin{algorithmic}[1]
\REQUIRE Unlabeled pool $\mathcal{U}$, initial labeled set $\mathcal{S}_0$, batch size $K$, balancing parameter $\lambda$, iterations $T$
\STATE Train initial model $f_{\theta_0}$ on $\mathcal{S}_0$
\FOR{$t = 1$ to $T$}
    \FOR{each $x \in \mathcal{U}$}
        \STATE Compute $\mathcal{H}(x)$ via Eq.~\eqref{eq:entropy}
        \STATE Generate $\text{CAM}_{\hat{y}}(x)$ via Grad-CAM
        \STATE Compute $D_{\text{exp}}(x)$ via Eq.~\eqref{eq:dice_misalignment}
        \STATE Compute $\text{Score}(x)$ via Eq.~\eqref{eq:composite_score}
    \ENDFOR
    \STATE $\mathcal{B}_t \leftarrow \text{TopK}_{x \in \mathcal{U}}(\text{Score}(x))$
    \STATE Retrieve labels and annotations for $\mathcal{B}_t$
    \STATE $\mathcal{S}_t \leftarrow \mathcal{S}_{t-1} \cup \mathcal{B}_t$
    \STATE Retrain $f_{\theta_t}$ on $\mathcal{S}_t$ using $\mathcal{L}_{\text{total}}$
    \STATE $\mathcal{U} \leftarrow \mathcal{U} \setminus \mathcal{B}_t$
\ENDFOR
\RETURN Final model $f_{\theta_T}$
\end{algorithmic}
\end{algorithm}

Algorithm~\ref{alg:active_learning} summarizes the complete procedure. As shown in Fig.~\ref{fig:active-learning-frm}, each iteration scores all unlabeled samples, selects the top-$K$, annotates them with labels and spatial ROIs, and retrains the model. The retraining step uses a composite loss~\cite{uddin2025expert}:
\begin{equation}
\mathcal{L}_{\text{total}} = \mathcal{L}_{\text{cls}} + \alpha \cdot \mathcal{L}_{\text{exp}},
\label{eq:total_loss}
\end{equation}
where $\mathcal{L}_{\text{cls}}$ is cross-entropy and $\mathcal{L}_{\text{exp}}$ is Dice loss between Grad-CAM and expert annotations. We set $\alpha = 0.10$ across all experiments. Crucially, as training progresses, the model's attention improves  which in turn makes $D_{\text{exp}}$ a more discriminative signal for subsequent rounds, creating a self-reinforcing cycle between better spatial reasoning and better sample selection.

\begin{figure*}[tbh]
    \centering
    \includegraphics[width=.48\linewidth]{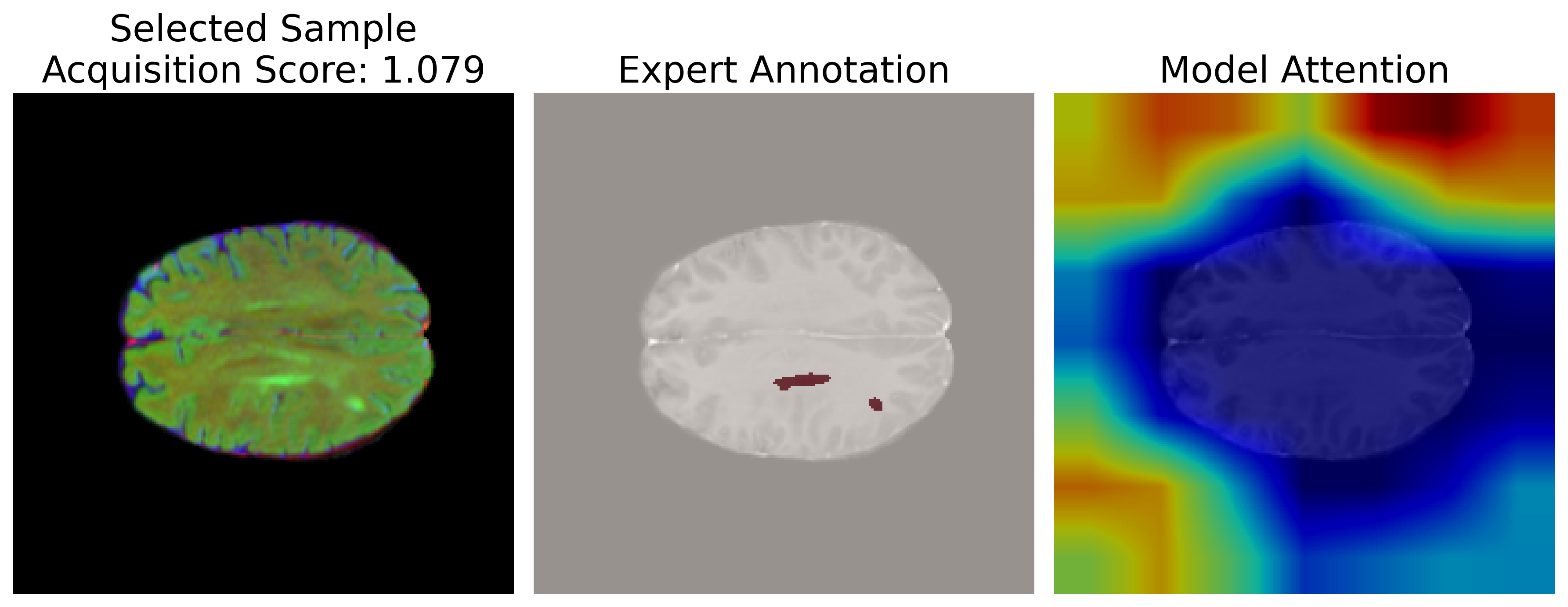}
    \includegraphics[width=.48\linewidth]{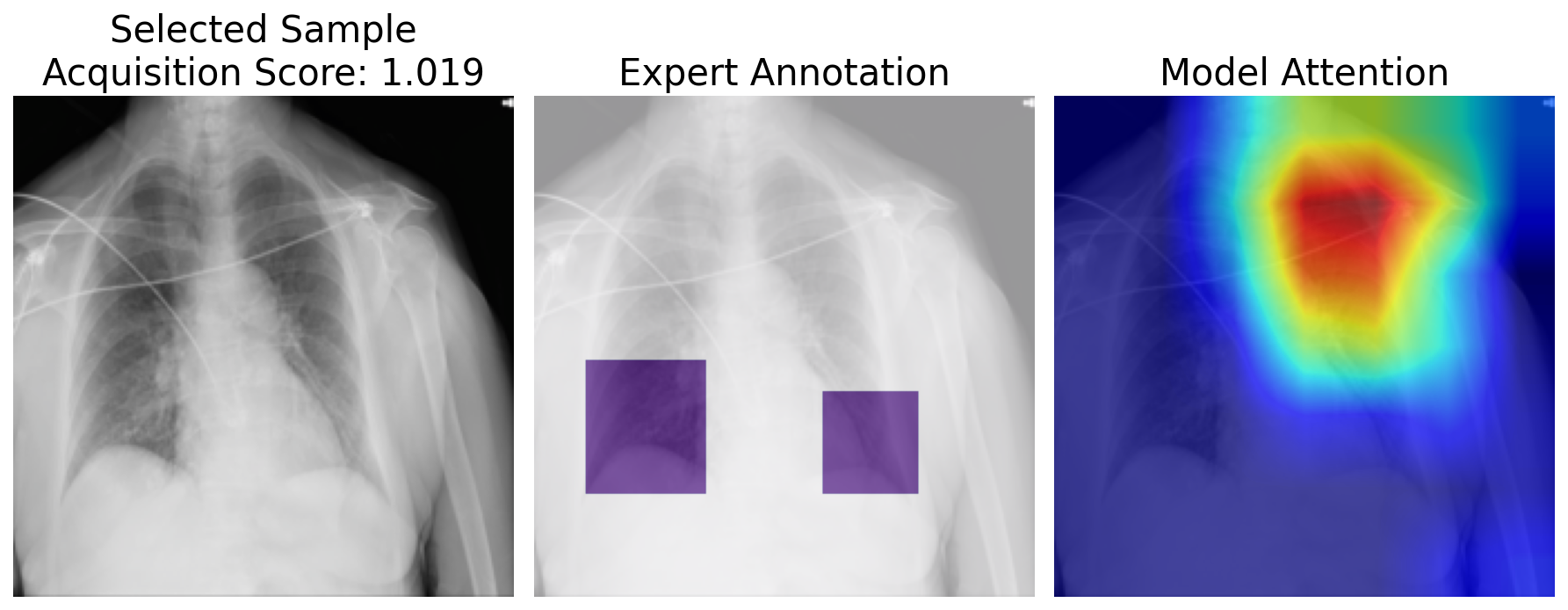}
    \includegraphics[width=.48\linewidth]{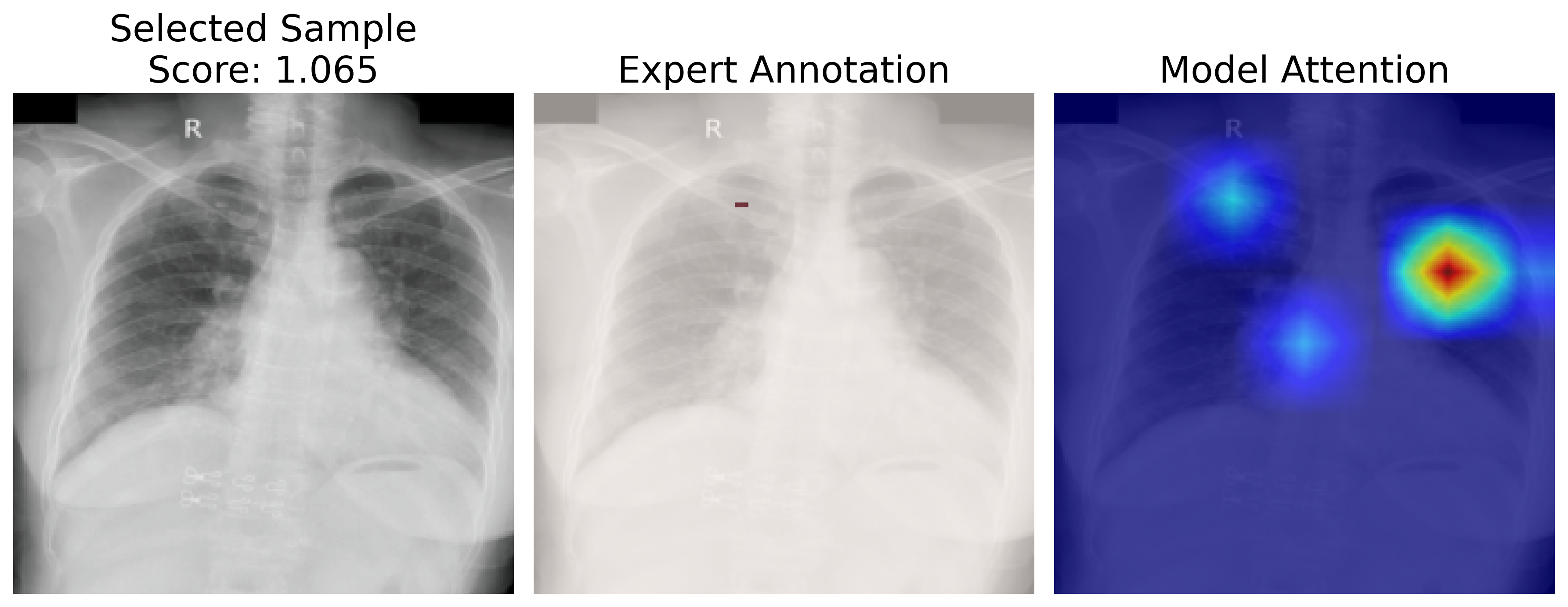}
    \includegraphics[width=.48\linewidth]{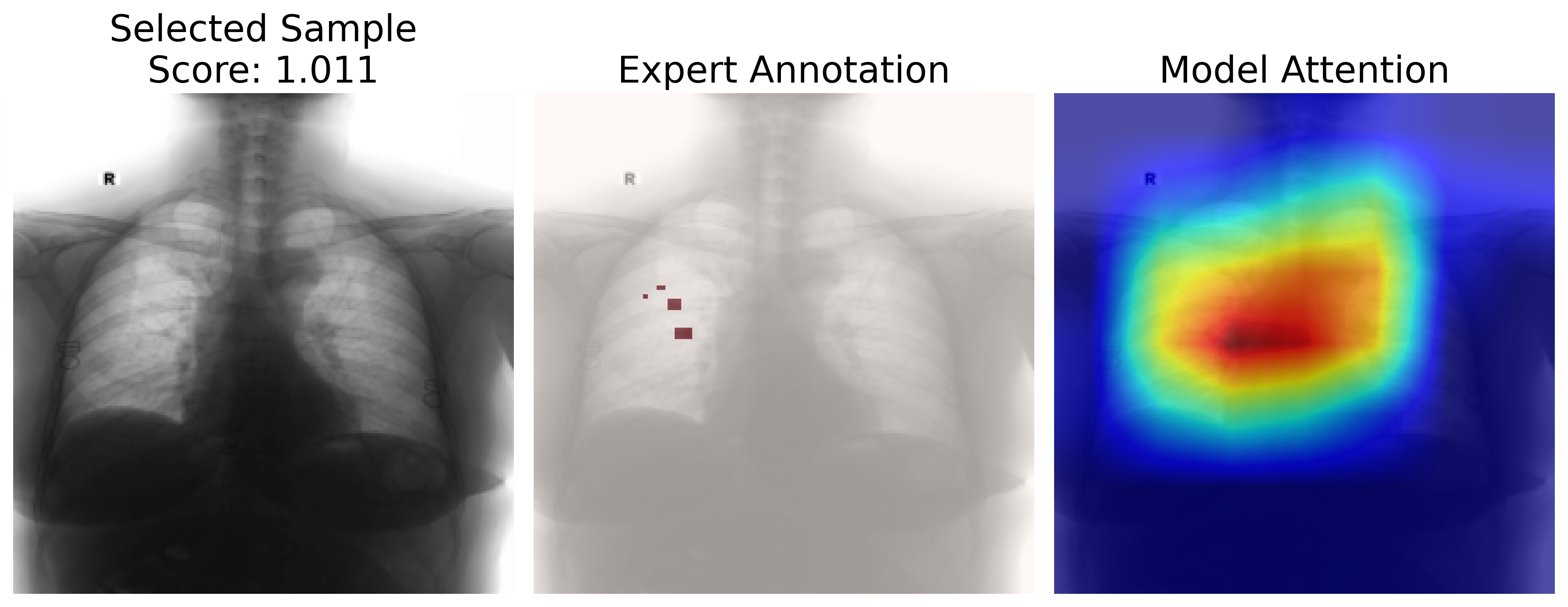}
    \caption{Representative high-scoring samples selected by EG-AL across different failure patterns. Each example shows: input image with acquisition score, expert annotation, and model attention. Top-left: BraTS case (score 1.079) with small tumor exhibiting high uncertainty and severe attention misalignment. Top-right: VinDr-CXR case (score 1.019) where model attention scatters across irrelevant regions. Bottom row: Two VinDr-CXR cases with poor spatial alignment despite varying confidence levels. Our dual-criterion scoring systematically identifies samples where models exhibit classification uncertainty, spatial misalignment, or both.}
    \label{fig:sample_selection}
\end{figure*}

\section{EXPERIMENTS}

\begin{table*}[tbh]
\centering
\caption{Performance comparison using 570 total samples over 7 AL rounds. Accuracy (\%) and macro AUC (\%) ± std over 5 seeds.}
\label{tab:active_learning_comparison}
\small
\renewcommand{\arraystretch}{1.2}
\begin{tabular}{|l|c|c|c|c|c|c|}
\hline
\multirow{2}{*}{\textbf{Dataset}} & \multicolumn{2}{c|}{\textbf{Baseline}} & \multicolumn{2}{c|}{\textbf{Random}} & \multicolumn{2}{c|}{\textbf{EG-AL}} \\
\cline{2-7}
& \textbf{Acc} & \textbf{Macro AUC} & \textbf{Acc} & \textbf{Macro AUC} & \textbf{Acc} & \textbf{Macro AUC} \\
\hline
\textbf{BraTS} & 45.10${\pm1.2}$ & 67.52${\pm1.5}$ & 58.01${\pm3.5}$ & 78.32${\pm2}$ & \textbf{77.22}${\pm1.4}$ & \textbf{90.00}${\pm1.1}$ \\
\hline
\textbf{VinDr-CXR} & 34.57${\pm3.2}$ & 56.39${\pm2.8}$ & 45.49${\pm3.5}$ & 58.21${\pm3.1}$ & \textbf{52.37}${\pm2.4}$ & \textbf{68.21}${\pm2.9}$ \\
\hline
\textbf{SIIM-COVID} & 30.32${\pm4.1}$ & 56.74${\pm3.7}$ & 38.28${\pm3.8}$ & 54.21${\pm4.2}$ & \textbf{52.66}${\pm2.7}$ & \textbf{66.92}${\pm3.3}$ \\
\hline
\end{tabular}
\end{table*}

\subsection{Experimental Setup}
We evaluate the proposed acquisition function on three expert-annotated 
medical imaging datasets  BraTS~\cite{menze2014multimodal}, 
VinDr-CXR~\cite{nguyen2022vindr}, and 
SIIM-FISABIO-RSNA-COVID-19~\cite{lakhani2021siim}  each providing 
diagnostic labels and spatial ROI annotations required for computing 
$D_{\text{exp}}$. We initialize EG-AL on 150 randomly selected samples 
(balanced across classes) using Eq.~\ref{eq:total_loss} with $\alpha=0.10$, 
then conduct seven AL rounds selecting 60 samples per round (570 total), 
comparing against random sampling under identical conditions. We use 
DenseNet-121~\cite{huang2017densely} pre-trained on ImageNet with a 
512-dimensional embedding layer, trained via prototypical 
networks~\cite{snell2017prototypical} with 5-shot episodes. To isolate 
the effect of $\lambda$ on acquisition quality, we perform grid search 
over $\{0.3, 0.5, 0.6, 0.7, 0.9\}$, setting $\lambda=0.5$ for BraTS 
and SIIM-COVID and $\lambda=0.6$ for VinDr-CXR; both extremes 
consistently underperform, confirming that neither criterion alone 
is sufficient as an acquisition signal. All experiments report mean 
accuracy and macro-averaged AUC over 5 random seeds with fixed test sets.

\begin{figure}[htbp]
    \centering
    \includegraphics[width=0.90\linewidth]{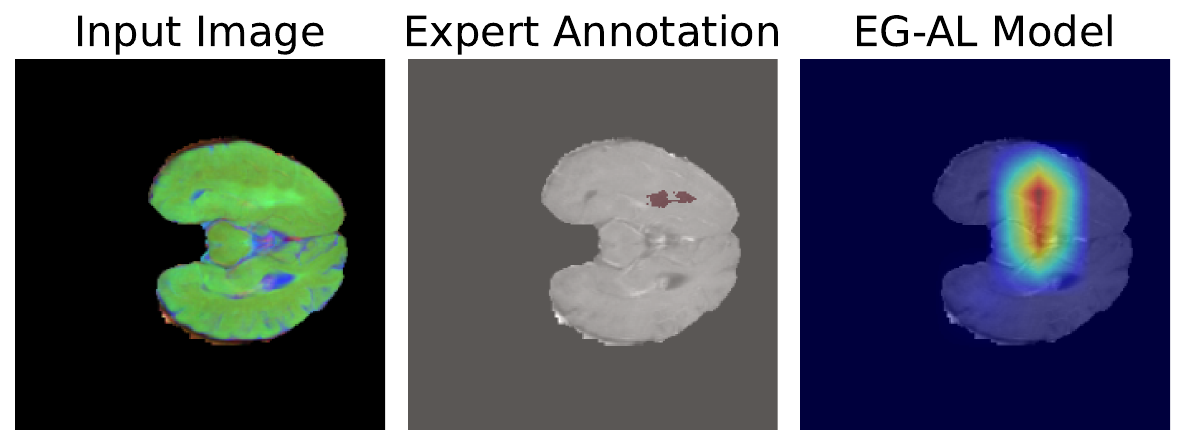}
    \includegraphics[width=0.90\linewidth]{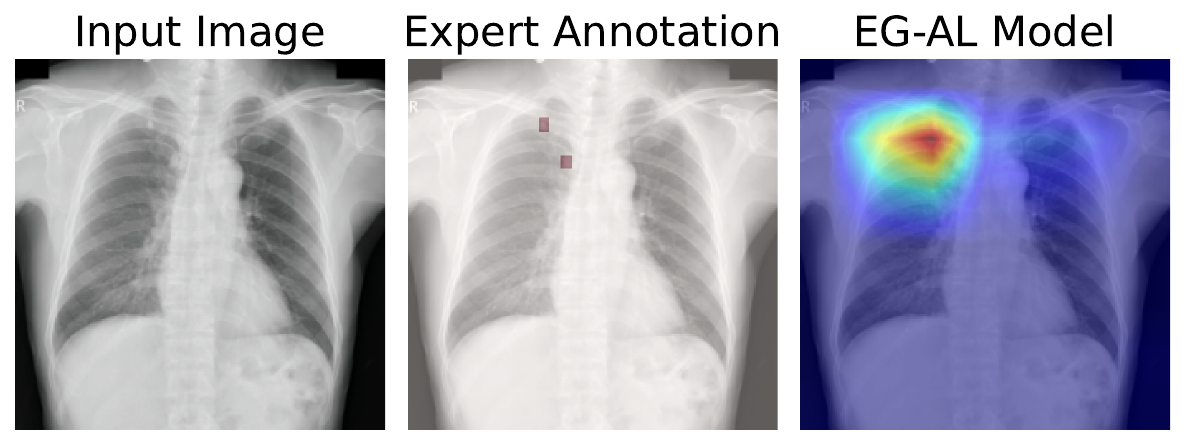}
    \includegraphics[width=0.90\linewidth]{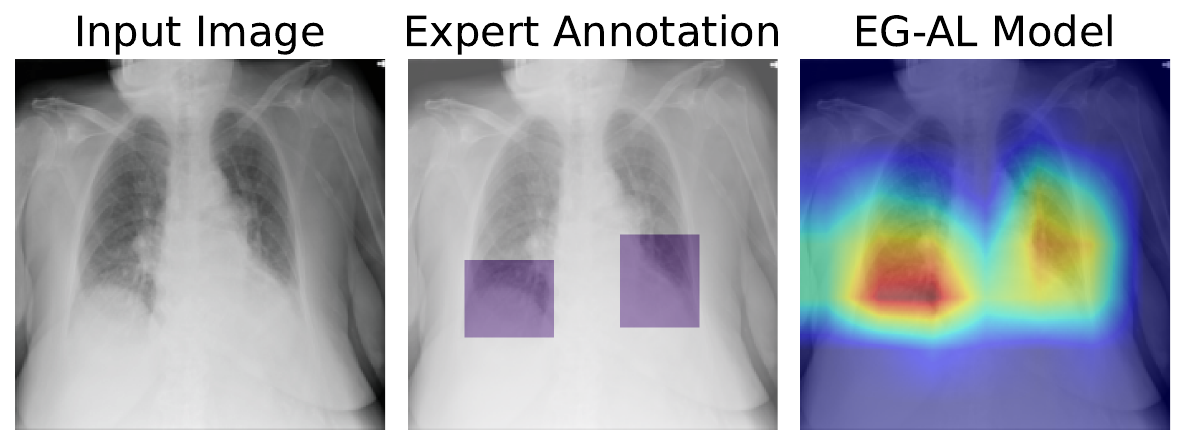}

 \caption{Attention alignment from models trained with EG-AL-selected 
    samples. Each row shows input image, expert annotation, and Grad-CAM 
    (left to right) for BraTS (top), VinDr-CXR (middle), and SIIM-COVID 
    (bottom). Models consistently localize expert-defined diagnostic regions, 
    validating that dual-criterion selection produces clinically aligned attention.}

    \label{fig:gradcam_comparison}
    \vspace{-5mm}
\end{figure}

\subsection{Performance Comparison}

Table~\ref{tab:active_learning_comparison} compares EG-AL against baseline 
and random sampling across all datasets, directly evaluating whether 
spatial misalignment is a valid and complementary acquisition signal to 
uncertainty.

For BraTS, EG-AL achieves 77.22\% accuracy and 90.00\% AUC, outperforming 
random sampling by 19.21\% and 18.68\% respectively. The margin confirms 
that selecting samples based on attention misalignment  not just label 
uncertainty  steers the model toward tumor boundaries it would otherwise 
learn to ignore.

On VinDr-CXR, EG-AL reaches 52.37\% accuracy and 68.21\% AUC, 
outperforming random sampling across both metrics. The gains here 
are particularly meaningful given the dataset's complexity: multiple 
overlapping thoracic findings make spurious attention especially 
likely, and $D_{\text{exp}}$ provides the signal to prioritize samples 
that correct it.

For SIIM-COVID, EG-AL achieves 52.66\% accuracy and 66.92\% AUC, 
substantially exceeding random sampling. COVID-19 severity classification 
requires precise localization of characteristic opacities  a spatial 
reasoning demand that uncertainty alone cannot enforce during acquisition.

Across all three datasets, EG-AL also exhibits lower standard deviation 
than random sampling, indicating that misalignment-aware acquisition 
produces more stable learning trajectories  a practically important 
property for clinical deployment.
improving both accuracy and clinical interpretability.

\subsection{Sample Selection Analysis}
Figure~\ref{fig:sample_selection} directly validates the three failure 
patterns identified by our acquisition function. The BraTS case 
(score 1.079) represents \textbf{high uncertainty with high misalignment}: 
the model both struggles to classify the small tumor and attends entirely 
to surrounding tissue, missing the expert-annotated boundary. The 
SIIM-COVID case (score 1.019) represents the clinically critical 
\textbf{low uncertainty, high misalignment} pattern: a confident 
prediction whose attention scatters across cardiac structures rather 
than the annotated lung opacities  a sample that any uncertainty-only 
criterion would discard. The VinDr-CXR cases capture \textbf{varying 
uncertainty with consistently poor spatial alignment} on subtle and 
distributed abnormalities. Together, these examples confirm that 
$D_{\text{exp}}$ recovers a distinct and complementary informativeness 
signal that entropy cannot provide.

\begin{figure}[tb]
    \centering
    \includegraphics[width=0.9\linewidth]{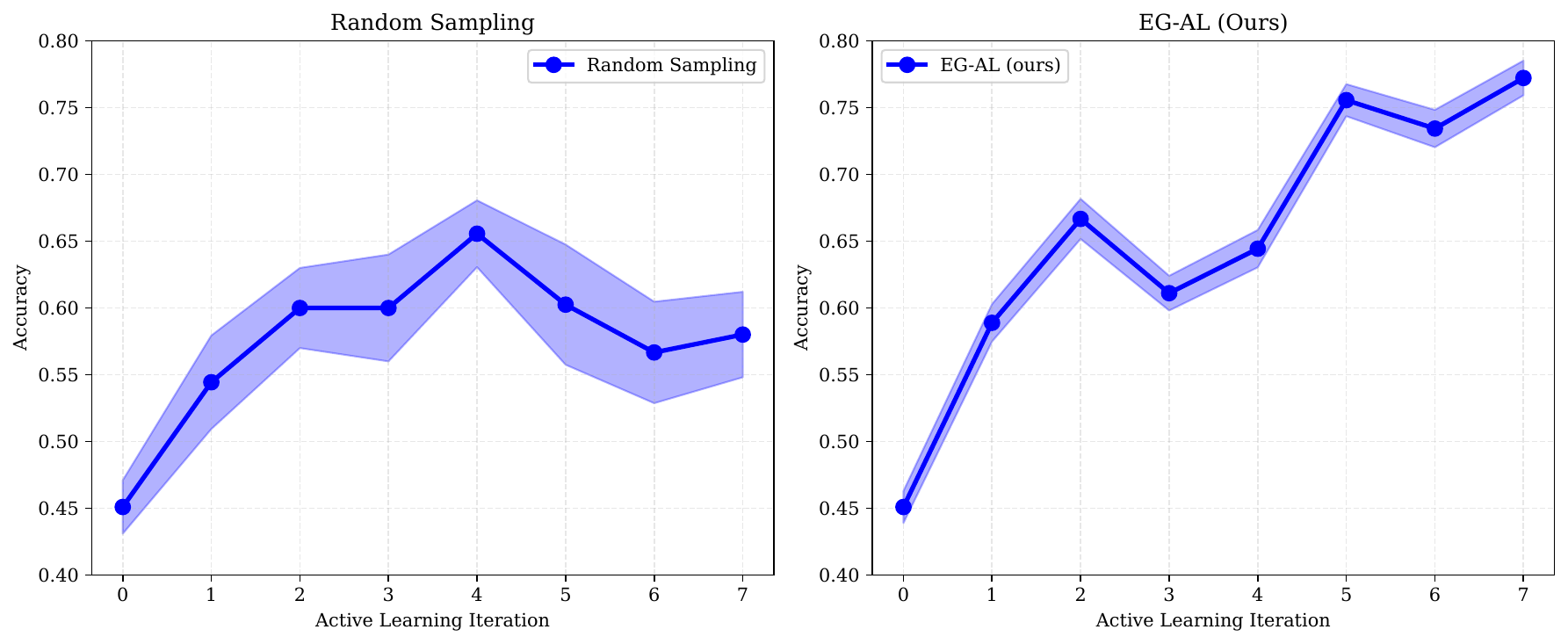}
    \includegraphics[width=0.9\linewidth]{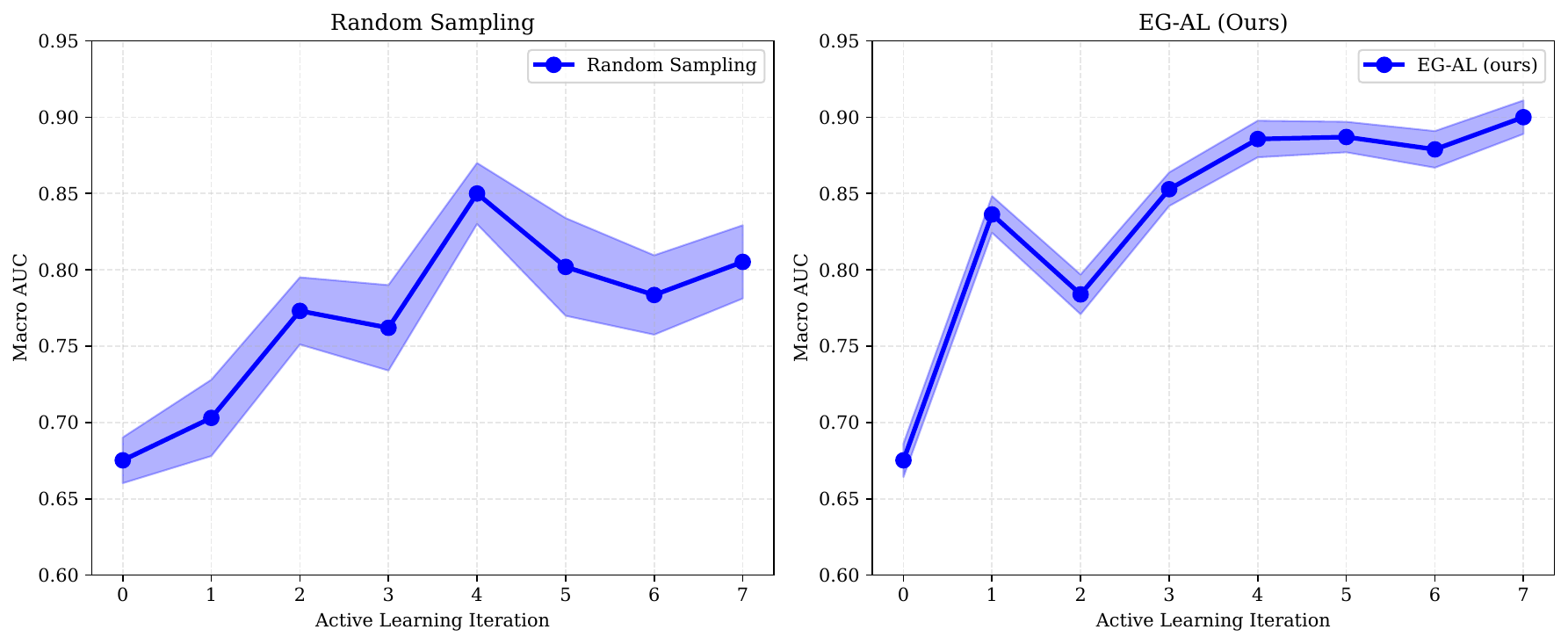}
    \caption{Progressive performance improvement on BraTS across active learning iterations. Top: Accuracy comparison. Bottom: Macro AUC comparison. Both methods start from the same baseline (iteration 0) and select 60 samples per round. EG-AL demonstrates consistent upward trajectory in both metrics, while random sampling exhibits higher fluctuation in both accuracy (top-left) and macro-auc (bottom-left). Shaded regions indicate standard deviation over 5 random seeds, showing EG-AL's superior stability.}
    \label{fig:al_progression}
    \vspace{-5mm}
\end{figure}

\subsection{Attention Alignment Visualization}
Figure~\ref{fig:gradcam_comparison} shows Grad-CAM attention from models  trained on EG-AL-selected samples across all three datasets. For BraTS, the model accurately localizes tumor boundaries in small and irregularly shaped cases. On VinDr-CXR, attention correctly highlights distributed thoracic abnormalities while avoiding spurious structures such as cardiac borders and ribs. For SIIM-COVID, the model consistently identifies characteristic lung opacities aligned with expert bounding boxes. These results confirm that selecting samples via $D_{\text{exp}}$ produces downstream models with improved spatial reasoning the acquisition criterion not only identifies spatial failures but actively corrects them.

\subsection{Active Learning Progression}
Figure~\ref{fig:al_progression} tracks EG-AL and random sampling across seven iterations on BraTS, both starting from the same baseline (45.1\% accuracy, 67.5\% AUC) and selecting 60 samples per round. Random sampling 
plateaus near 60\% accuracy with high variance; EG-AL reaches 77.2\% on a consistent upward trajectory  a 19.2\% absolute gain. The AUC progression is similarly decisive: EG-AL reaches 90\% with tight confidence bands versus random sampling's 80.5\% with substantially higher variance. The widening gap across iterations demonstrates that 
misalignment is not just a useful signal at initialization  it becomes increasingly discriminative as the model improves, consistent with the self-reinforcing cycle described earlier in methodology section.

\section{CONCLUSION}

We presented EG-AL, a dual-criterion acquisition function that selects samples for annotation based on both classification uncertainty and spatial attention misalignment with expert ROIs. Our central finding 
is that $D_{\text{exp}}$  Dice-based divergence between Grad-CAM attention and expert annotations  constitutes a valid and complementary acquisition signal to entropy, recovering a class of informative samples (confident predictions with spatially wrong attention) that uncertainty-only methods cannot detect. Across three medical imaging datasets and two 
modalities, EG-AL consistently outperforms random sampling with only 570 labeled samples, and ablation confirms that neither criterion alone achieves the same data efficiency.
This work demonstrates that the question of \textit{which samples to annotate} should be answered not only in terms of label uncertainty, but also in terms of spatial reasoning quality  opening a new axis along which active learning acquisition functions can be designed and evaluated in safety-critical domains.

 \section*{Acknowledgment}
This work was supported by the National Science Foundation under Grant No. \href{https://www.nsf.gov/awardsearch/showAward?AWD_ID=2346643}{\#2346643}, the U.S. Department of Defense under Award No. \href{https://dtic.dimensions.ai/details/grant/grant.14525543}{\#FA9550-23-1-0495}, and the U.S. Department of Education under Grant No. P116Z240151.
Any opinions, findings, conclusions or recommendations expressed in this material are those of the author(s) and do not necessarily reflect the views of the National Science Foundation, the U.S. Department of Defense, or the U.S. Department of Education.






\bibliographystyle{IEEEtran}
\bibliography{ref}

@inproceedings{he2016deep,
 title={Deep residual learning for image recognition},
author={He, Kaiming and Zhang, Xiangyu and Ren, Shaoqing and Sun, Jian},
booktitle={Proceedings of the IEEE conference on computer vision and pattern recognition},
pages={770--778},
year={2016}
}

@article{snell2017prototypical,
title={Prototypical networks for few-shot learning},
author={Snell, Jake and Swersky, Kevin and Zemel, Richard},
journal={Advances in neural information processing systems},
volume={30},
year={2017}
}

@misc{ren2021surveydeepactivelearning,
title={A Survey of Deep Active Learning}, 
author={Pengzhen Ren and Yun Xiao and Xiaojun Chang and Po-Yao Huang and Zhihui Li and Brij B. Gupta and Xiaojiang Chen and Xin Wang},
year={2021},
eprint={2009.00236},
archivePrefix={arXiv},
primaryClass={cs.LG}, 
}

@inproceedings{aghdam2019active,
title={Active learning for deep detection neural networks},
author={Aghdam, Hamed H and Gonzalez-Garcia, Abel and Weijer, Joost van de and L{\'o}pez, Antonio M},
booktitle={Proceedings of the IEEE/CVF International Conference on Computer Vision},
pages={3672--3680},
year={2019}
}

@ARTICLE{wang2017cost,
author={Wang, Keze and Zhang, Dongyu and Li, Ya and Zhang, Ruimao and Lin, Liang},
journal={IEEE Transactions on Circuits and Systems for Video Technology}, 
title={Cost-Effective Active Learning for Deep Image Classification}, 
year={2017},
volume={27},
number={12},
pages={2591-2600},
doi={10.1109/TCSVT.2016.2589879}}

@article{selvaraju2017grad,
title={Grad-cam: Visual explanations from deep networks via gradient-based localization},
author={Selvaraju, Ramprasaath R and Cogswell, Michael and Das, Abhishek and Vedantam, Ramakrishna and Parikh, Devi and Batra, Dhruv},
booktitle={Proceedings of the IEEE international conference on computer vision},
pages={618--626},
year={2017} 
 }

@article{lundberg2017unified,
title={A unified approach to interpreting model predictions},
author={Lundberg, Scott M and Lee, Su-In},
journal={Advances in neural information processing systems},
volume={30},
year={2017}
}

@article{menze2014multimodal,
title={The multimodal brain tumor image segmentation benchmark (BRATS)},
author={Menze, Bjoern H and Jakab, Andras and Bauer, Stefan and Kalpathy-Cramer, Jayashree and Farahani, Keyvan and Kirby, Justin and Burren, Yuliya and Porz, Nicole and Slotboom, Johannes and Wiest, Roland and others},
journal={IEEE transactions on medical imaging},
volume={34},
number={10},
pages={1993--2024},
year={2014},
publisher={IEEE}
}

@article{nguyen2022vindr,
title={VinDr-CXR: An open dataset of chest X-rays with radiologist's annotations},
author={Nguyen, Ha Q and Lam, Khanh and Le, Linh T and Pham, Hieu H and Tran, Dat Q and Nguyen, Dung B and Le, Dung D and Pham, Chi M and Tong, Hang TT and Dinh, Diep H and others},
journal={Scientific Data},
volume={9},
number={1},
pages={429},
year={2022},
publisher={Nature Publishing Group UK London}
}

@misc{lakhani2021siim,
author = {Lakhani, Paras and Mongan, John and Singhal, Chinmay and Zhou, Quan and Andriole, Katherine P. and Auffermann, William F. and Prasanna, Prasanth and Pham, Thuy and Peterson, Michael and Bergquist, Peter J. and Cook, Thomas S. and Ferraciolli, Sergio F. and de Antonio Corradi, Gabriel C. and Takahashi, Masahiro and Workman, Sean S. and Parekh, Mayur and Kamel, Safwan and Galant, James H. and Mas-Sanchez, Alvaro and Benítez, Edgar C. and Sánchez-Valverde, Miguel and Jaques, Lukas and Panadero, Manuel and Vidal, Manel and Culiáñez-Casas, Miquel and Angulo-Gonzalez, Diego M. and Langer, Sergio G. and de la Iglesia Vaya, María and Shih, George},
title = {The 2021 SIIM-FISABIO-RSNA Machine Learning COVID-19 Challenge: Annotation and Standard Exam Classification of COVID-19 Chest Radiographs},
year = {2021},
howpublished = {OSF Preprints},
month = {October 21},
doi = {10.31219/osf.io/532ek}
}

@inproceedings{hsu2015active,
title={Active learning by learning},
author={Hsu, Wei-Ning and Lin, Hsuan-Tien},
booktitle={Proceedings of the AAAI Conference on Artificial Intelligence},
volume={29},
number={1},
year={2015}
}

@article{ho2024learning,
title={Learning to learn for few-shot continual active learning},
author={Ho, Stella and Liu, Ming and Gao, Shang and Gao, Longxiang},
journal={Artificial Intelligence Review},
volume={57},
number={10},
pages={280},
year={2024},
publisher={Springer}
}

@incollection{biswas2023active,
title={Active learning on medical image},
author={Biswas, Angona and Abdullah Al, Nasim Md and Ali, Md Shahin and Hossain, Ismail and Ullah, Md Azim and Talukder, Sajedul},
booktitle={Data Driven Approaches on Medical Imaging},
pages={51--67},
year={2023},
publisher={Springer}
}

@article{gaillochet2023active,
title={Active learning for medical image segmentation with stochastic batches},
author={Gaillochet, M{\'e}lanie and Desrosiers, Christian and Lombaert, Herv{\'e}},
journal={Medical Image Analysis},
volume={90},
pages={102958},
year={2023},
publisher={Elsevier}
}

@article{zhou2021active,
title={Active, continual fine tuning of convolutional neural networks for reducing annotation efforts},
author={Zhou, Zongwei and Shin, Jae Y and Gurudu, Suryakanth R and Gotway, Michael B and Liang, Jianming},
journal={Medical image analysis},
volume={71},
pages={101997},
year={2021},
publisher={Elsevier}
}

@article{hao2021transfer,
title={A transfer learning--based active learning framework for brain tumor classification},
author={Hao, Ruqian and Namdar, Khashayar and Liu, Lin and Khalvati, Farzad},
journal={Frontiers in artificial intelligence},
volume={4},
pages={635766},
year={2021},
publisher={Frontiers Media SA}
}

@article{loh2022application,
title={Application of explainable artificial intelligence for healthcare: A systematic review of the last decade (2011--2022)},
author={Loh, Hui Wen and Ooi, Chui Ping and Seoni, Silvia and Barua, Prabal Datta and Molinari, Filippo and Acharya, U Rajendra},
journal={Computer methods and programs in biomedicine},
volume={226},
pages={107161},
year={2022},
publisher={Elsevier}
}

@article{sadeghi2024review,
title={A review of Explainable Artificial Intelligence in healthcare},
author={Sadeghi, Zahra and Alizadehsani, Roohallah and Cifci, Mehmet Akif and Kausar, Samina and Rehman, Rizwan and Mahanta, Priyakshi and Bora, Pranjal Kumar and Almasri, Ammar and Alkhawaldeh, Rami S and Hussain, Sadiq and others},
journal={Computers and Electrical Engineering},
volume={118},
pages={109370},
year={2024},
publisher={Elsevier}
}

@article{vsefvcik2023improving,
title={Improving a neural network model by explanation-guided training for glioma classification based on MRI data},
author={{\v{S}}ef{\v{c}}{\'\i}k, Franti{\v{s}}ek and Benesova, Wanda},
journal={International Journal of Information Technology},
volume={15},
number={5},
pages={2593--2601},
year={2023},
publisher={Springer}
}

@inproceedings{sun2021explanation,
title={Explanation-guided training for cross-domain few-shot classification},
author={Sun, Jiamei and Lapuschkin, Sebastian and Samek, Wojciech and Zhao, Yunqing and Cheung, Ngai-Man and Binder, Alexander},
booktitle={2020 25th international conference on pattern recognition (ICPR)},
pages={7609--7616},
year={2021},
organization={IEEE}
}

@inproceedings{ravi2017optimization,
title={Optimization as a model for few-shot learning},
author={Ravi, Sachin and Larochelle, Hugo},
booktitle={International conference on learning representations},
year={2017}
}

@article{caragliano2025doctorintheloopexplainablemultiviewdeep,
title={Doctor-in-the-Loop: An explainable, multi-view deep learning framework for predicting pathological response in non-small cell lung cancer},
author={Caragliano, Alice Natalina and Ruffini, Filippo and Greco, Carlo and Ippolito, Edy and Fiore, Michele and Tacconi, Claudia and Nibid, Lorenzo and Perrone, Giuseppe and Ramella, Sara and Soda, Paolo and others},
journal={Image and Vision Computing},
pages={105630},
year={2025},
publisher={Elsevier}
}

@article{zhang2025uncertainty,
title={An uncertainty-incorporated active data diffusion learning framework for few-shot equipment RUL prediction},
author={Zhang, Chao and Gong, Daqing and Xue, Gang},
journal={Reliability Engineering \& System Safety},
volume={254},
pages={110632},
year={2025},
publisher={Elsevier}
}

@article{zhan2025new,
title={A new active learning surrogate model for time-and space-dependent system reliability analysis},
author={Zhan, Hongyou and Xiao, Ning-Cong},
journal={Reliability Engineering \& System Safety},
volume={253},
pages={110536},
year={2025},
publisher={Elsevier}
}

@article{chen2025survey,
title={A survey on deep learning in medical image registration: New technologies, uncertainty, evaluation metrics, and beyond},
author={Chen, Junyu and Liu, Yihao and Wei, Shuwen and Bian, Zhangxing and Subramanian, Shalini and Carass, Aaron and Prince, Jerry L and Du, Yong},
journal={Medical Image Analysis},
volume={100},
pages={103385},
year={2025},
publisher={Elsevier}
}

@inproceedings{uddin2025expert,
title={Expert-Guided Explainable Few-Shot Learning for Medical Image Diagnosis},
author={Uddin, Ifrat Ikhtear and Wang, Longwei and Santosh, KC},
booktitle={Data Engineering in Medical Imaging: Third MICCAI Workshop, DEMI 2025, Held in Conjunction with MICCAI 2025, Daejeon, South Korea, September 27, 2025, Proceedings},
pages={95},
year={2025},
organization={Springer Nature}
}

@inproceedings{huang2017densely,
title={Densely connected convolutional networks},
author={Huang, Gao and Liu, Zhuang and Van Der Maaten, Laurens and Weinberger, Kilian Q},
booktitle={Proceedings of the IEEE conference on computer vision and pattern recognition},
pages={4700--4708},
year={2017}
}

@article{wang2019representation,
title={Representation learning and nature encoded fusion for heterogeneous sensor networks},
author={Wang, Longwei and Liang, Qilian},
journal={IEEE Access},
volume={7},
pages={39227--39235},
year={2019},
publisher={IEEE}
}

@inproceedings{wang2014congestion,
title={Congestion aware dynamic user association in heterogeneous cellular network: A stochastic decision approach},
author={Wang, Longwei and Chen, Wen and Li, Jun},
booktitle={2014 IEEE International Conference on Communications (ICC)},
pages={2636--2640},
year={2014},
organization={IEEE}
}

@inproceedings{wang2024enhanced,
title={Enhanced robustness by symmetry enforcement},
author={Wang, Longwei and Ghimire, Aashish and Santosh, K and Zhang, Zheng and Li, Xueqian},
booktitle={IEEE Conference on Artificial Intelligence (IEEE CAI) 2024},
year={2024}
}

@article{wang2021explaining,
title={Explaining the behavior of neuron activations in deep neural networks},
author={Wang, Longwei and Wang, Chengfei and Li, Yupeng and Wang, Rui},
journal={Ad Hoc Networks},
volume={111},
pages={102346},
year={2021},
publisher={Elsevier}
}

@inproceedings{shi2019deep,
title={Deep reinforcement learning based computation offloading for mobility-aware edge computing},
author={Shi, Minyan and Wang, Rui and Liu, Erwu and Xu, Zhixin and Wang, Longwei},
booktitle={International conference on communications and networking in china},
pages={53--65},
year={2019},
organization={Springer International Publishing Cham}
}

@article{wang2021improving,
title={Improving robustness of deep neural networks via large-difference transformation},
author={Wang, Longwei and Wang, Chengfei and Li, Yupeng and Wang, Rui},
journal={Neurocomputing},
volume={450},
pages={411--419},
year={2021},
publisher={Elsevier}
}

@inproceedings{xiao2022looking,
title={Looking Beyond Content: Modeling and Detection of Fake News from a Social Context Perspective.},
author={Xiao, Kenan and Wang, Longwei and Gupta, Ashish and Qin, Xiao},
booktitle={HICSS},
pages={1--10},
year={2022}
}

@inproceedings{wang2019layer,
title={Layer-wise entropy analysis and visualization of neurons activation},
author={Wang, Longwei and Chen, Peijie and Wang, Chengfei and Wang, Rui},
booktitle={International Conference on Communications and Networking in China},
pages={29--36},
year={2019},
organization={Springer International Publishing Cham}
}

@article{wang2024dense,
title={Dense cross-connected ensemble convolutional neural networks for enhanced model robustness},
author={Wang, Longwei and Li, Xueqian and Zhang, Zheng},
journal={arXiv preprint arXiv:2412.07022},
year={2024}
}

@article{wall2025winsor,
title={Winsor-CAM: Human-tunable visual explanations from deep networks via layer-wise Winsorization},
author={Wall, Casey and Wang, Longwei and Rizk, Rodrigue and Santosh, KC},
journal={arXiv preprint arXiv:2507.10846},
year={2025}
}

@inproceedings{wang2025explainability,
title={Explainability-driven defense: grad-CAM-guided model refinement against adversarial threats},
author={Wang, Longwei and Uddin, Ifrat Ikhtear and Qin, Xiao and Zhou, Yang and Santosh, KC},
booktitle={Proceedings of the AAAI Symposium Series (AAAI) 2025},
volume={6},
number={1},
pages={49--57},
year={2025}
}

@inproceedings{ranabhat2025multi,
title={Multi-scale unrectified push-pull with channel attention for enhanced corruption robustness},
author={Ranabhat, Robin Narsingh and Wang, Longwei and Qin, Xiao and Zhou, Yang and Santosh, KC},
booktitle={Proceedings of the AAAI Symposium Series 2025},
volume={6},
number={1},
pages={34--41},
year={2025}
}

@inproceedings{wang2025bridging,
title={Bridging Symmetry and Robustness: On the Role of Equivariance in Enhancing Adversarial Robustness},
author={Wang, Longwei and Uddin, Ifrat Ikhtear and Santosh, KC and Zhang, Chaowei and Qin, Xiao and Zhou, Yang},
booktitle={Advances in Neural Information Processing Systems (NeurIPS) 2025},
year={2025}
}

@INPROCEEDINGS{10928239,
  author={Sabuj, Md Sanowar Hossain and Imam, Touhid and Islam, Jahirul and Sultana, Sharmin and Uddin, Mohammad Shihab and Uddin, Bushra},
  booktitle={2024 IEEE International Conference on Computing (ICOCO)},
  title={Recondite Thyroid Pathology Prediction: Hermeneutic Integration of Neural and Machine Learning Architectures},
  year={2024},
  volume={},
  number={},
  pages={267-272},

  doi={10.1109/ICOCO62848.2024.10928239}}

@INPROCEEDINGS{10928056,
  author={Uddin, Bushra and Uddin, Mohammad Shihab and Sultana, Sharmin and Sabuj, MD Sanowar Hossain and Neha, Fariha Ferdous and Uddin, MD Salah},
  booktitle={2024 International Conference on Computer and Applications (ICCA)},
  title={Epistemological Advancements in Cardiological Forecasting: Machine Learning as a Paradigm for Prognostic Precision},
  year={2024},
  volume={},
  number={},
  pages={01-06},
  doi={10.1109/ICCA62237.2024.10928056}}

@INPROCEEDINGS{10927825,
  author={Sultana, Sharmin and Uddin, Bushra and Uddin, Mohammad Shihab and Uddin, MD Salah and Karim, Fazle and Mehedi, Mohiuddin},
  booktitle={2024 International Conference on Computer and Applications (ICCA)},
  title={Analyzing Neuroimaging Epiphenomena: Machine Learning Approaches in Alzheimer's Prognostication},
  year={2024},
  volume={},
  number={},
  pages={1-6},
  doi={10.1109/ICCA62237.2024.10927825}}

\end{document}